\documentclass[twocolumn,amsmath,amssymb]{revtex4}
\usepackage{graphicx}
\usepackage{dcolumn}
\usepackage{bm}
\usepackage{caption}
\usepackage{subcaption}
\usepackage{amsmath}
\begin{document}

\title{Topological resonance in Weyl semimetals in circularly-polarized optical pulse}

\author{Fatemeh Nematollahi}
\author{S. Azar Oliaei Motlagh}
\author{Jhih-Sheng Wu}
\author{Rupesh Ghimire}
\author{Vadym Apalkov}
\author{Mark I. Stockman}
\affiliation{Center for Nano-Optics (CeNO) and
Department of Physics and Astronomy, Georgia State
University, Atlanta, Georgia 30303, USA
}

\date{\today}
\begin{abstract}
We study theoretically the ultrafast electron dynamics of three-dimensional Weyl semimetals in the field of a laser pulse. For a circularly-polarized pulse, such 
dynamics is governed by topological resonance, which manifests itself as a specific  conduction band population distribution in the vicinity of the Weyl points. The topological resonance is determined by the competition between the topological phase and the dynamic phase and depends on the handedness of a circularly polarized 
pulse.  Also, we show that the conduction band population induced by a circularly-polarized pulse that consists of two oscillations with opposite handedness is highly chiral, which represents the intrinsic chirality of the Weyl points.
  \end{abstract}
\maketitle

\section{Introduction}

The interaction of ultrafast laser pulses with solids was a subject of intensive theoretical and experimental research 
over the last two decades. Such interaction is characterized by highly non-linear electron dynamics and strong perturbation of  electron systems and can be used to probe and control the transport and optical properties of solids within a femtosecond time scale \cite{wachter2014ab,nematollahi2018phosphorene,
su2017ultrafast,nematollahi2019weyl_spie,
motlagh2018femtosecond,motlagh2019ultrafast}. 
The ultrafast electron dynamics in the field of the pulse is determined by electron energy dispersion and interband dipole matrix elements. Such matrix elements strongly depend on the topology of the system and have singularities in the reciprocal space for topologically nontrivial solids. 
Such singularities strongly modify ultrafast electron dynamics and result in unique 
features in both 
electron population of the conduction band and generated electric currents. One of the materials with 
nontrivial topology is a 2D graphene monolayer, which belongs to the class of 
Dirac semimetals that have linear energy dispersion near special points called the Dirac points \cite{young2012dirac}.

Graphene is a unique material, which has amazing electrical, optical, and mechanical properties as well as numerous potential applications~\cite{koppens2014photodetectors,novoselov2012roadmap}. Graphene is a single layer of carbon atoms with a honeycomb crystal structure that has two inequivalent Dirac points ($\textbf{K}$ and $\textbf{K}^\prime$) in the first Brillouin zone. These two Dirac points have non-zero Berry flux of 
opposite signs, which makes graphene a locally topologically nontrivial material, while the total Berry flux through the whole Brillouin zone is zero. 
While in graphene, the relativistic 
energy dispersion and non-zero Berry flux at the Dirac points are realized in 2D, the corresponding Dirac points can be also 
realized in 3D solids, i.e., in Dirac semimetals. In 3D, the Dirac points are double
degenerate. Such degeneracy is protected by time-reversal and inversion 
symmetries of the solid. To lift the degeneracy, either time-reversal symmetry $\mathcal{T}$ \cite{burkov2011weyl} or inversion symmetry $\mathcal{P}$  \cite{halasz2012time} should be broken.
In this case the Dirac point is transformed into a set of separated Weyl points, which are monopoles 
of Berry curvature. Such materials are called Weyl semimetals. 
Due to the fermion doubling theorem \cite{nielsen1981absence}, the Weyl points appear in pairs with opposite chiralities, i.e., the Weyl points in a pair are sink or source of the Berry curvature. 

Recent studies have revealed that Weyl semimetals show strong nonlinear optical response such as the second harmonic generation (SHG) \cite{zyuzin2017chiral} and nonlinear Hall effect~\cite{sodemann2015quantum}. These nonlinear effects are of topological origin and are due to large Berry curvature localized near the Weyl points. Furthermore, a circularly polarized pulse can excite electrons near the Weyl points selectively \cite{ma2017direct,yu2016determining}. Such selectivity can open new opportunities for device  applications.

The response of the Weyl semimetals to a linearly polarized ultrafast pulse has been studied theoretically in Ref. \cite{nematollahi2019weyl}. The results show that the electron dynamics in such materials is coherent and highly anisotropic. 
At the same time, in Ref. \cite{motlagh2018femtosecond,motlagh2019topological} it was shown that the response of a solid to a circularly polarized ultrashort pulse can have some extra features. Namely, 
for gapped graphene-like materials, such as transition metal dichalcogenides, the electrons experience topological resonance in the field of circularly polarized pulse, while there is no such resonance for linearly polarized pulse. The topological resonance occurs due to competition between the topological phase and the dynamic phase and results in predominant population of one of the valleys of 
graphene-like materials. The topological resonance occurs only for gapped 
graphene materials but not for pristine graphene and strongly depends on the 
magnitude of the band gap. In this relation the Weyl semimental becomes an 
interesting system to study the topological resonance, since near each Weyl point, in the reciprocal space, the 
2D cross sections are equivalent to gapped graphene systems with the band gap that 
depends of the distance of the 2D cross section to the Weyl point. Thus by studying the interaction of Weyl semimentals with circularly polarized pulse we can study the ultrafast electron dynamics for both pristine graphene and gapped graphene 
with different band gaps. In this paper we consider the interaction of Weyl semimetals with ultrafast circularly polarized pulses of different handedness. We identify the  features of topological resonance in the conduction band population distribution in the reciprocal space of Weyl semimetal.

\section{Model and Equations}\label{II}
 
\begin{figure}
\begin{center}\includegraphics[width=0.45\textwidth]{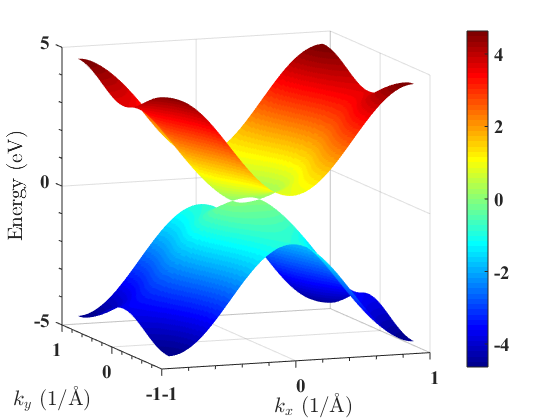}\end{center}
\caption{(Color online) Energy dispersion of TaAs as a function of $k_x$ and $k_y$ at $k_z=0$. }
\label{Figure_Energy}
\end{figure}

In the presence of an external electric field, the full Hamiltonian of the system becomes 
\begin{equation}
H(t)=H_0+e\textbf{F}(t)\textbf{r}, 
\end{equation}
where $e$  is the electron charge and  $H_0$ is the field-free Hamiltonian of the system. For the field-free 
Hamiltonian we use the two-band model of  Weyl semimetals. Such 
Hamiltonian $H_0$ describes the low-energy excitations near two Weyl points 
located at $k_w^{\pm}=(\pm{k_0},0,0)$ in the reciprocal space. The Hamiltonian has the following form \cite{hasan2017discovery}
\begin{equation}
{\cal H}_0=A(\textbf{k})\sigma_x+B(\textbf{k})\sigma_y+C(\textbf{k})\sigma_z,\  \label{Hamiltonian}
\end{equation}
where $\textbf{k}=(k_x,k_y,k_z)$ is a vector of the reciprocal space, $\sigma_x$, $\sigma_y$, $\sigma_z$ are Pauli 
matrices, and $A(\textbf{k})$, $B(\textbf{k})$, $C(\textbf{k})$ are given by the following expressions
\begin{eqnarray}
A(\textbf{k})& = &t_x(\cos(k_xa)-\cos(k_0a))+ \nonumber\\
& & t_y(\cos(k_yb)-1)+t_z(\cos(k_zc)-1)\nonumber,\\ 
B(\textbf{k}) & = & t_y\sin(k_yb)\nonumber,\\
C(\textbf{k}) & = & t_z\sin(k_zc).
\end{eqnarray}
Here $a$, $b$ and $c$ are lattice constants along $x$, $y$ and $z$ directions, respectively, and $t_x$, $t_y$, $t_z$ are hopping integrals which are related to the Fermi velocities $v_x$, $v_y$, and $v_z$ at the Weyl points through the following expressions 
\begin{align}
 v_x&=-(a/\hbar) t_x\sin(\pm{k_0a}),\\ \nonumber
 v_y&=(b/\hbar) t_y,\\ \nonumber
 v_z&=(c/\hbar) t_z. \
\label{system1}
\end{align}
 We apply our analysis to TaAs Weyl semimetal, which has a body-centered tetragonal lattice system, with lattice constants $a=b=3.437$ $\text{\AA}$ along $x$ and $y$ directions, respectively and $c=11.646$ $\text{\AA}$ along $z$ direction. The space group in TaAs is $I4_1 md$ ($\#109$, $C_{4v}$) \cite{murray1976phase}. The important symmetries are the time-reversal symmetry ($\mathcal{T}$), the four-fold rotational symmetry around the $\hat{z}$ axis ($C_{4z}$) and two mirror reflections about $x=0$ and $y=0$. TaAs has $24$ Weyl points: eight Weyl points are located at $(\pm0.0072 \pi/a,0.4827 \pi/b,1 .000\pi/c)$ and are called $W_1$ and sixteen Weyl points are located at $(\pm0.0185 \pi/a,0.2831 \pi/b,0.6000\pi/c)$ and are called $W_2$ \cite{lv2015experimental,lee2015fermi}. The two band Hamiltonian (\ref{Hamiltonian}) is used to describe the electron dynamics near one pair of Weyl points. Below we consider the Weyl points that are located at $(\pm0.1,0,0)$. The hopping integrals are $t_x=1.8801$ eV, $t_y=0.4917$, eV and $t_z =0.1646$ eV. 
The energy dispersion of TaAs near the Weyl points as a function of $k_x$ and $k_y$ and at $k_z=0$ 1/\AA~ is shown in Fig.\ref{Figure_Energy}.
 
We assume that during the pulse the electron dynamics is coherent and is  
 described by the time-dependent Schr\"odinger equation (TDSE)  
\begin{equation}
i\hbar\frac{d\Psi}{dt}=\hat{H}(t)\Psi.
\label{shrodinger}     
\end{equation} 
The electric field of the pulse generates both interband and intraband electron dynamics. The intraband dynamics is determined by the Bloch acceleration theorem \cite{bloch1929quantenmechanik}
\begin{equation}
\hbar\frac{d\textbf{k}}{dt}=e\textbf{F}(t),      
\end{equation}  
solution of which has the following form
\begin{equation}
\textbf{k}(\textbf{q},t)=\textbf{q}+\frac{e}{\hbar}\int_{-\infty}^{t}\textbf{F}(t^{\prime})dt^{\prime},      
\end{equation}
where $\textbf{q}$ is the initial electron wave vector.

 The corresponding wave functions, which are solutions of the time-dependent Schr\"odinger equation within 
a single band, are Houston functions \cite{houston1940acceleration} and are given by the following expression  
\begin{equation}
\Phi_{\alpha\textbf{q}}^{(H)}(\textbf{r},t)=\Psi_{\textbf{k}(\textbf{q},t)}^{(\alpha)}(\textbf{r})\exp{\bigg(i\phi_{\alpha}^{(D)}(\textbf{q},t)+i\phi_{\alpha}^{(B)}(\textbf{q},t)\bigg)},
\end{equation} 
where $\Psi_{\textbf{k}}^{(\alpha)}(\textbf{r})$ are Bloch wave functions with wave vector $\textbf{k}$ at 
band $\alpha$, $\alpha =v$ for the valence band and $\alpha=c$ for the conduction band. The dynamic phase, $\phi_{\alpha}^{(D)}$, and the geometric (Berry) phase, $\phi_{\alpha}^{(B)}$, are defined by the following expressions
\begin{equation}
\phi_{\alpha}^{(D)}(\textbf{q},t)={-\frac{1}{\hbar}\int_{-\infty}^{t}dt^{\prime}E_\alpha[\textbf{k}(\textbf{q},t^{\prime})]},
\end{equation}
\begin{equation}
\phi_{\alpha}^{(B)}(\textbf{q},t)={\frac{e}{\hbar}\int_{-\infty}^{t}dt^{\prime} \textbf{F}(t^{\prime})\textbf{A}^{\alpha \alpha}[\textbf{k}(\textbf{q},t^{\prime})]}, 
\end{equation}
where $\textbf{A}^{\alpha \alpha}=\langle \Psi_{\textbf{q}}^{(\alpha)}|i\frac{\partial}{\partial \textbf{q}}|\Psi_{\textbf{q}}^{(\alpha)}\rangle$ is the intraband Berry connection for band $\alpha$.

It is convenient to express the solution of  TDSE (\ref{shrodinger}) in the basis of Houston functions as 
\begin{equation}
\Psi_{\textbf{k}(\textbf{q},t)}^{(\alpha)}(\textbf{r})=\sum_{\alpha=v,c} \beta_{\alpha \textbf{q}}(t)\Phi_{\alpha \textbf{q}}^{(H)}(\textbf{r},t),  
\end{equation}
where $\beta_{\alpha \textbf{q}}(t)$ are expansion coefficients. These coefficients satisfy the following 
system of equations
\begin{equation}
i\hbar\frac{\partial\textbf{\textit{B}}_\textbf{q}(t)}{\partial{t}}=H^\prime(\textbf{q},t)\textbf{\textit{B}}_\textbf{q}(t),
\label{Hamiltonian_interaction}
\end{equation} 
where $\textbf{\textit{B}}_\textbf{q}(t)$ and Hamiltonian $H^\prime(\textbf{q},t)$ are defined as 
\begin{equation}
\textbf{\textit{B}}_\textbf{q}(t)=
\begin{pmatrix}
\beta_{v,\textbf{q}}(t)\\
\beta_{c,\textbf{q}}(t)
\end{pmatrix},
\end{equation}
and
\begin{equation}
H^\prime(\textbf{q},t)=e\textbf{F}(t)\hat{\textbf{A}}(\textbf{q},t),
\end{equation}  
where
\begin{equation}
\hat{\textbf{A}}(\textbf{q},t)=\begin{bmatrix}
0 & \mathcal{D}^{cv}(\textbf{q},t)\\
\mathcal{D}^{vc}(\textbf{q},t)& 0
\end{bmatrix},
\end{equation}
\begin{eqnarray}
\mathcal{D}^{cv}(\textbf{q},t)&=&\textbf{A}^{cv}[\textbf{k}(\textbf{q},t)]\times \\\nonumber
&&\exp\bigg({i\big[\phi_{cv}^{(B)}(\textbf{q},t)+\phi_{cv}^{(D)}(\textbf{q},t)\big]}\bigg),
\end{eqnarray}
\begin{equation}
\phi_{cv}^{(\mathrm{D})}(\textbf{q},t)=\phi_{v}^{(\mathrm{D})}(\textbf{q},t)-\phi_{c}^{(\mathrm{D})}(\textbf{q},t)
\label{dynamic_phase}   
\end{equation}
\begin{equation}
\phi_{cv}^{(\mathrm{B})}(\textbf{q},t)=\phi_{v}^{(\mathrm{B})}(\textbf{q},t)-\phi_{c}^{(\mathrm{B})}(\textbf{q},t)
\label{dynamic_phase}   
\end{equation}
Here, $\phi_{cv}^{(\mathrm{D})}(\textbf{q},t)$ is a dynamic phase,
 $\phi_{cv}^{(\mathrm{B})}(t)$ is a topological phase,  and 
 matrix $\emph{\textbf{A}}_{cv}(\textbf{k})$ is the non-Abelian $k$-space gauge potential called Berry connection and expressed as \cite{berry1984quantal,xiao2010berry}
\begin{equation}
\textbf{A}^{cv}(\textbf{q})=\langle\Psi_\textbf{q}^{(c)}|i\frac{\partial}{\partial \textbf{q}}|\Psi_\textbf{q}^{(v)}\rangle,
\label{Dipole} 
\end{equation}
which can be found analytically in the case of Weyl semimetals  as
\begin{eqnarray}
\lefteqn{\textbf{A}_x^{vc}=(\textbf{A}_x^{cv})^{*}=\frac{t_x a}{ 2i\bigg(A^2(\textbf{k})+B^2(\textbf{k})+C^2(\textbf{k})\bigg)}} \\\nonumber
&&\times \frac{\sin(k_xa)}{\sqrt{A^2(\textbf{k})+B^2(\textbf{k})}}\\\nonumber
&&\times \bigg[A(\textbf{k})C(\textbf{k})-iB(\textbf{k})\sqrt{A^2(\textbf{k})+B^2(\textbf{k})+C^2(\textbf{k})}\bigg],\\
\lefteqn{\textbf{A}_y^{vc}=(\textbf{A}_y^{cv})^{*}=\frac{t_y a}{ 2i\bigg(A^2(\textbf{k})+B^2(\textbf{k})+C^2(\textbf{k})\bigg)}}\\\nonumber
&&\times \frac{1}{\sqrt{A^2(\textbf{k})+B^2(\textbf{k})}}\Bigg[C(\textbf{k})\bigg(A(\textbf{k})\sin(k_ya)-\\\nonumber
&&B(\textbf{k})\cos(k_ya)\bigg)-i\sqrt{A^2(\textbf{k})+B^2(\textbf{k})+C^2(\textbf{k})}\\\nonumber
&&\times \bigg(A(\textbf{k})\cos(k_ya)+B(\textbf{k})\sin(k_ya)\bigg)\Bigg],\\
\lefteqn{\textbf{A}_z^{vc}=(\textbf{A}_z^{cv})^{*}=\frac{t_zc\sqrt{A^2(\textbf{k})+B^2(\textbf{k})}}{2i\bigg(A^2(\textbf{k})+B^2(\textbf{k})+C^2(\textbf{k})\bigg)}}\\\nonumber
&&\times\bigg[\cos(k_zc)+\frac{\sin(k_zc)}{A^2(\textbf{k})+B^2(\textbf{k})}\\\nonumber
&&\times\bigg(A(\textbf{k})C(\textbf{k})-iB(\textbf{k})\sqrt{A^2(\textbf{k})+B^2(\textbf{k})+C^2(\textbf{k})}\bigg)\bigg].
\label{velocities}
\end{eqnarray}
We numerically solve differential equation (\ref{Hamiltonian_interaction}) with initial conditions  $(\beta_{v\textbf{q}},\beta_{c\textbf{q}})=(1,0)$. 
A general solution can be expressed in terms of the evolution operator $\hat{U}(\textbf{q},t)$ as 
\begin{equation}
\textbf{\textit{B}}_{\textbf{q}}(t)=\hat{U}(\textbf{q},t)\textbf{\textit{B}}_{\textbf{q}}(-\infty),
\label{evolution_operator} 
\end{equation}  
\begin{equation}
\hat{U}(\textbf{q},t)=\hat{T}\exp [i\int_{t^\prime=-\infty}^{t}\hat{\textbf{A}}(\textbf{q},t^\prime)d\textbf{k}(\textbf{q},t^\prime)],
\label{evolution_operator}
\end{equation}
where $\hat{T}$ denotes the time-ordering operator and the integral is affected along the Bloch trajectory $\textbf{k}(\textbf{q},t)$. We characterize the electron dynamics in terms of the conduction band 
population distribution $N_{CB}(\textbf{q},t)=|\beta_{cq}(t)|^2$ in the reciprocal space.

\section{Results and Discussion}
 The rest of the paper is organized as follows: In the subsection (A) we consider a circularly polarized pulse propagating along $z$ direction. We characterize the electron dynamics in the field of the pulse by the residual conduction band (CB) population distribution in the reciprocal space. The data show that a circularly polarized single oscillation pulse induces the topological resonance in the system. In the next subsection (B) we apply a circularly polarized pulse that consists of two cycles. The optical pulse propagates in $z$ direction and induces a CB population distribution in the reciprocal space, which is highly chiral and shows the chirality of the Weyl points.
 
\subsection{A single oscillation circularly-polarized pulse} 
\begin{figure}
\begin{center}\includegraphics[width=0.5\textwidth]{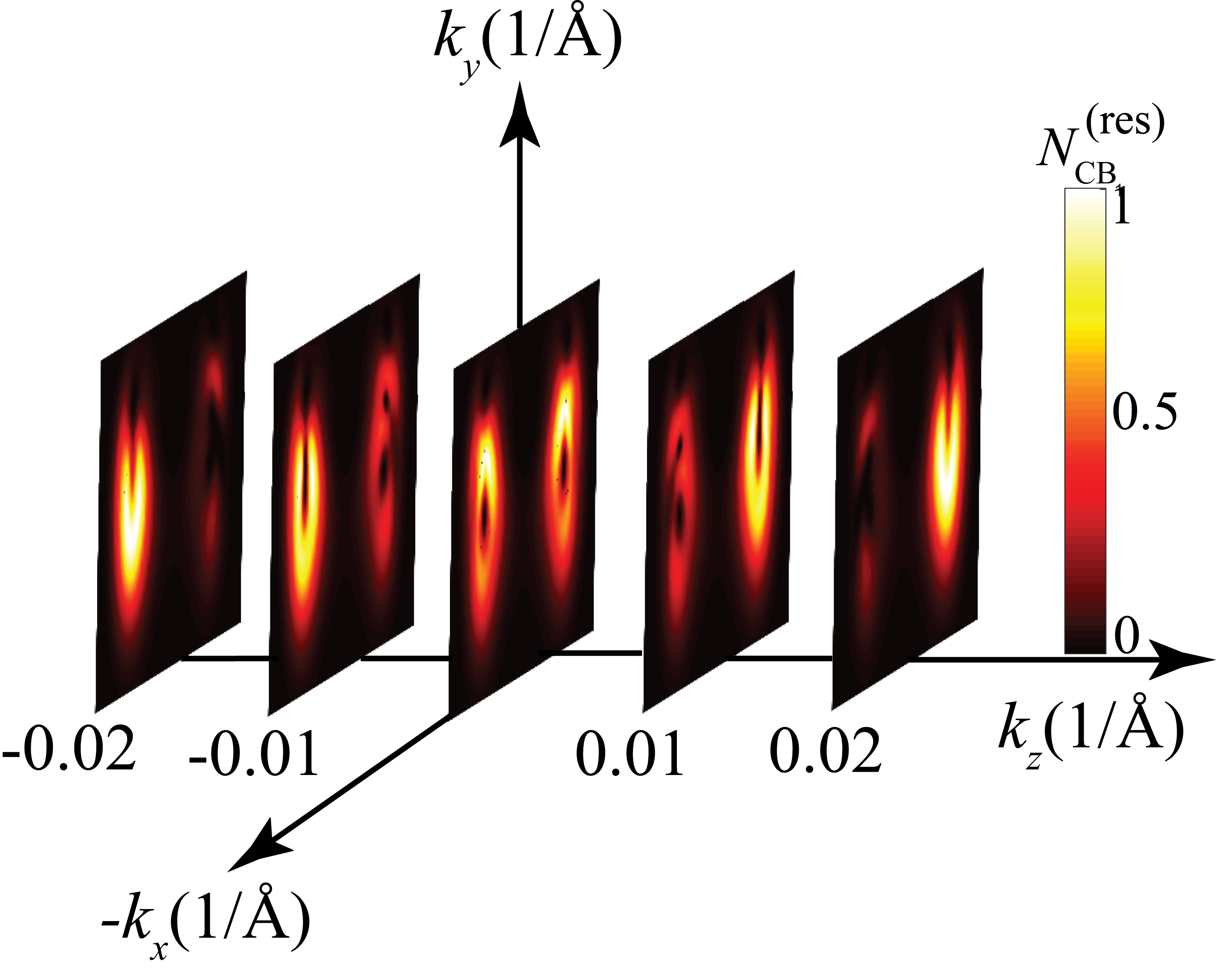}\end{center}
\vspace*{-0.25cm}
\caption{(Color online) Residual CB population distribution as a function of $(k_x,k_y)$ for different values of $k_z$  after a single oscillation left-handed circularly polarized pulse. The pulse with the field amplitude of $F_0=3$~ mV/\AA~ propagates along the $z$ direction.}
\label{Figure_1_pulse}
\end{figure}

  We study the response of Weyl semimetals to a single-oscillation circularly polarized pulse in terms of residual CB population in the reciprocal space. We assume that the left-handed circularly polarized optical pulse propagates along $z$-direction and its components are  defined as
\begin{align}
 F_x(t)&=-F_0e^{-u^2}(1-2u^2),& F_y(t)&=2uF_0e^{-u^2}, 
\label{system1}
\end{align}
where the field amplitude $F_0=3$ mV/\AA, $u=t/\tau$ and $\tau =10$ fs is the pulse duration. We numerically solve TDSE, see Eq. (\ref{Hamiltonian_interaction}), with  initial condition  $(\beta_{v\textbf{q}},\beta_{c\textbf{q}})=(1,0)$. Applied optical field causes redistribution of electrons between the valence and conduction bands, which results in finite CB population. 
 The residual CB population, i.e., the CB population at the end of the pulse, $t=25$ fs, is shown in Fig.\ref{Figure_1_pulse}.
 
\begin{figure}
\begin{center}\includegraphics[width=0.5\textwidth]{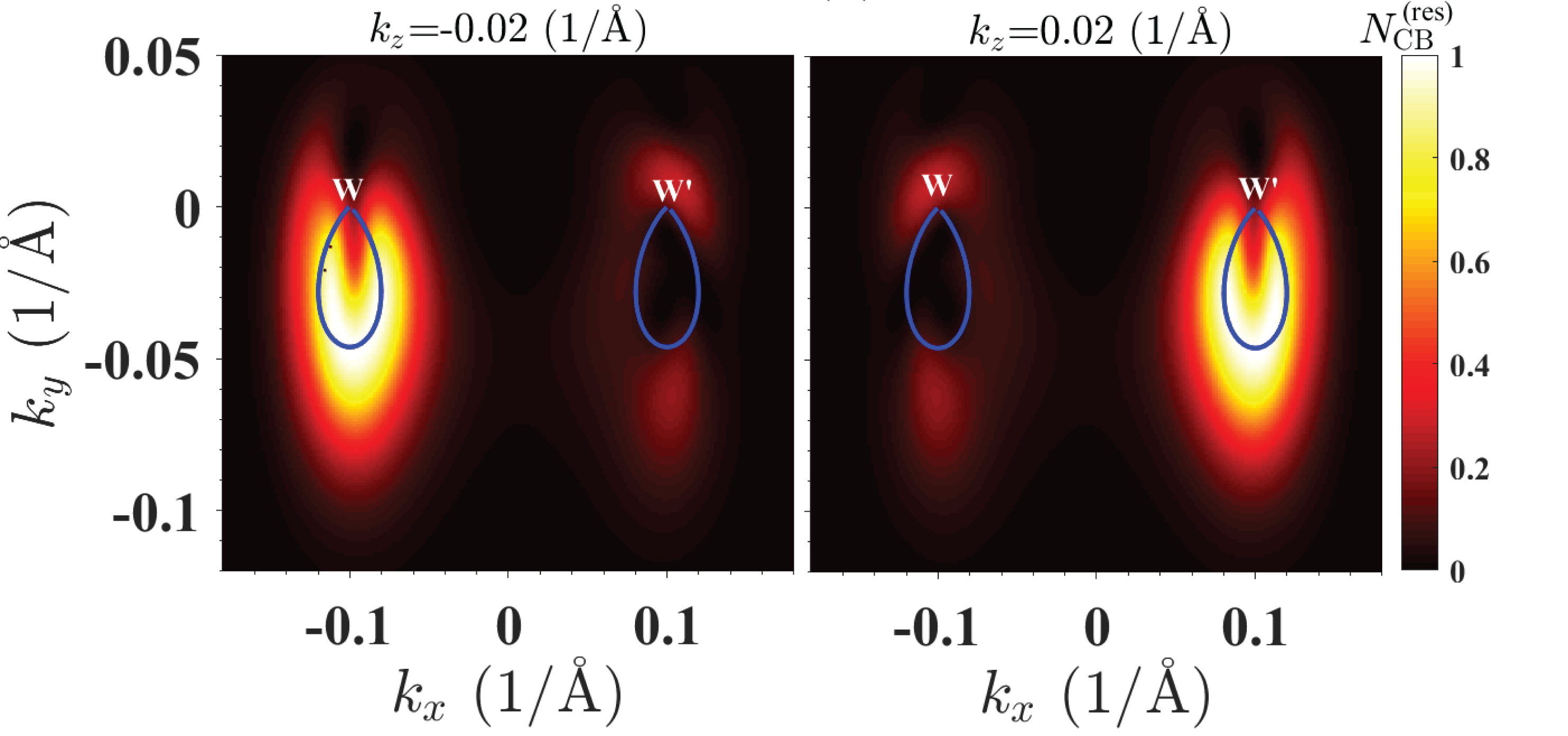}\end{center}
\caption{Residual CB population distribution in the reciprocal space as a function of $(k_x,k_y)$ for  $k_z=\pm 0.2$ (1/\AA) after a single oscillation right-handed circularly polarized pulse. The pulse with the field amplitude $F_0=3$ mV/\AA~ propagates along the $z$ direction. The separatrix is shown by a solid blue line.}
\label{Figure_cw}
\end{figure}

 At $k_z=0$, the electron system in $(k_x,k_y)$ plane is equivalent to pristine 
 graphene and the responses at both Weyl points to an external electric field are similar. The corresponding CB population distribution is symmetric with respect to $y$-axes. The large CB population is located near the Weyl points, see Fig. \ref{Figure_1_pulse},  which correlates with the profile of the interband dipole coupling that is the strongest near the Weyl points.  This property is similar to what we have in graphene \cite{kelardeh2016attosecond} and is due to singularity of the dipole coupling exactly at the Weyl points or Dirac points for graphene. Since the interband dipole matrix elements are highly localized near the Weyl points, the conduction band population distribution has sharp maxima along the separatrix (solid blue line in Fig.\ref{Figure_cw}). Here the separatrix is defined as a set 
of points in the reciprocal space, for which the electron trajectory during the pulse goes directly through the Weyl point.

For a non-zero value of $k_z$, the electron system in the $k_x-k_y$ plane, near each Weyl point, becomes similar to gapped graphene \cite{motlagh2019topological} with the band gap that is proportional to $|k_z|$. The corresponding CB population distribution as a function of $k_x$ and $k_y$ and for different $k_z$ is shown in Fig. \ref{Figure_1_pulse}. The data show that for non-zero $k_z$, the $W$, $(-0.1,0,0)$, and $W^{'}$, $(0.1,0,0)$, points are populated differently. For $k_z>0$, the CB is highly populated in the vicinity of $W^{'}$ point, while it is less populated for $k_z<0$. This is different from the CB population distribution for $k_z=0$ and is due to the fact that for $k_z\neq 0$, the effective 2D system (in $k_x$-$k_y$ plane) becomes similar to gapped graphene, for which there is the effect of topological resonance. The origin of the topological resonance can be understood by looking at the expression for the CB population in the first order of the perturbation theory. Namely, within this approximation, the CB population is given by the following expression
\begin{equation}
n_{CB}=\Bigg|\oint|\mathcal{A}_{cv}[\textbf{k}(\textbf{q},t)]\textbf{n}(t)|\exp\bigg(i\phi_{cv}^{\text{(tot)}}(\textbf{q},t)\bigg)dk(\textbf{q},t)\Bigg|^2,
\label{residual}
\end{equation}
where $\textbf{n}(t)=\textbf{F}(t)/F(t)$ is the unit vector tangential to the Bloch trajectory and the total phase $\phi^{(\text{tot})}$ is 
\begin{equation}
\phi_{cv}^{\text{(tot)}}=\phi_{cv}^{\text{(B)}}(\textbf{q},t)+\phi_{cv}^{\text{(A)}}(\textbf{q},t)+\phi_{cv}^{\text{(D)}}(\textbf{q},t),
\label{phase_total}
\end{equation}    
  where the dipole matrix element phase, $\phi_{cv}^{\text{(A)}}(\textbf{q},t)$ is defined as 
\begin{equation}
\phi_{cv}^{\text{(A)}}(\textbf{q},t)=\arg\bigg({\mathcal{A}_{cv}(\textbf{k}(\textbf{q},t)\textbf{n}(t)}\bigg).
\end{equation}
Since $\bigg|\textbf{A}_{cv}[\textbf{k}(\textbf{q},t)]\textbf{n}(t)\bigg|$ is a smooth function of time, the residual CB population, Eq. (\ref{residual}), is determined by oscillating phase factor $\exp[i\phi_{cv}^{(tot)}(\textbf{q},t)]$ and the topological resonance occurs when the total phase is stationary, i.e., the topological phase, which is the combination of the geometric phase and the phase
of the dipole matrix element, and the dynamic phase cancel each other.

\begin{figure}
\begin{center}\includegraphics[width=0.5\textwidth]{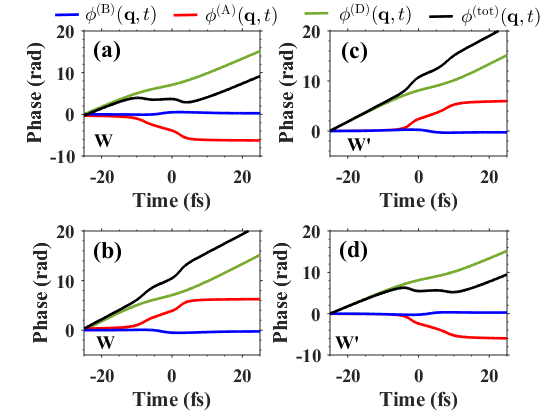}\end{center}
\vspace*{-0.25cm}
\caption{(Color online) Phases $\phi_{cv}^{(tot)}(\textbf{q},t)$,
$\phi_{cv}^{(B)}(\textbf{q},t)$, $\phi_{cv}^{(A)}(\textbf{q},t)$, and $\phi_{cv}^{(D)}(\textbf{q},t)$ for different initial wave vectors in the vicinity of the separatrix. The initial wave vectors, $(q_x,q_y,q_z)$, are (a) $(-0.07,-0.06,-0.02)$ (b) $(-0.07,-0.06,0.02)$ .(c) $(0.07,-0.06,-0.02)$ (d) $(0.07,-0.06,0.02)$. The amplitude of the pulse is $F_0=3$ mV/\AA~ and it is right-hand circularly polarized.  }
\label{Figure_phase}
\end{figure}

Figure \ref{Figure_cw} shows the residual CB population distribution at $k_z=\pm0.02$ 1/\AA ~induced by a circularly polarized pulse.  The topological resonance, which manifests itself as a large CB population, can be explained by  
  Fig. \ref{Figure_phase}, where the corresponding phases, $\phi_{cv}^{\text{(A)}}(\textbf{q},t)$, $\phi_{cv}^{\text{(B)}}(\textbf{q},t)$, $\phi_{cv}^{\text{(D)}}(\textbf{q},t)$ and $\phi_{cv}^{\text{(tot)}}(\textbf{q},t)$ are shown. 
  
  Since the magnitude of the interband coupling is the strongest near the Weyl points, which corresponds to $t=0$ in Fig. \ref{Figure_phase}, then we need to study the behavior of the total phase at $t$ close to zero.
  For a given Weyl point, the phases $\phi_{cv}^{\text{(A)}}(\textbf{q},t)$ and $\phi_{cv}^{\text{(B)}}(\textbf{q},t)$, have opposite signs. Figures \ref{Figure_phase}(a) and (b) show the different phases for a point near the $W$ point for $k_z=-0.02$ 1/\AA ~and $k_z=0.02$, 1/\AA ~ respectively. The topological resonance is determined by behavior of the total phase, $\phi_{cv}^{\text{(tot)}}(\textbf{q},t)$, around $t=0$ fs. The total phase is almost constant for $k_z=-0.02$ 1/\AA~ , and has strong time dependent for  $k_z=0.02$ 1/\AA~. Thus, the topological resonance results in large CB population for the $W$ point at $k_z=-0.02$ 1/\AA~ while the CB population is relatively small for the $W$ point at $k_z=0.02$ 1/\AA~ plane. Opposite, for the $W^\prime$ point, the total phase is almost constant around $t=0$~fs for $k_z=0.02$ 1/\AA, which results in corresponding large CB population, see Figs. \ref{Figure_phase} (c)-(d).
  
The response of Weyl semimetals to an ultrafast circularly polarized optical pulse is the same as what was predicted for gapped graphene. In the case of gapped graphene, the right-hand circularly polarized pulse mostly populates 
the $K$ valley while the CB population at the $K^{'}$ valley is small \cite{motlagh2019topological}. 

\subsection{Two-cycle circularly-polarized pulse}

\begin{figure}
\begin{center}
\includegraphics[width=80mm]{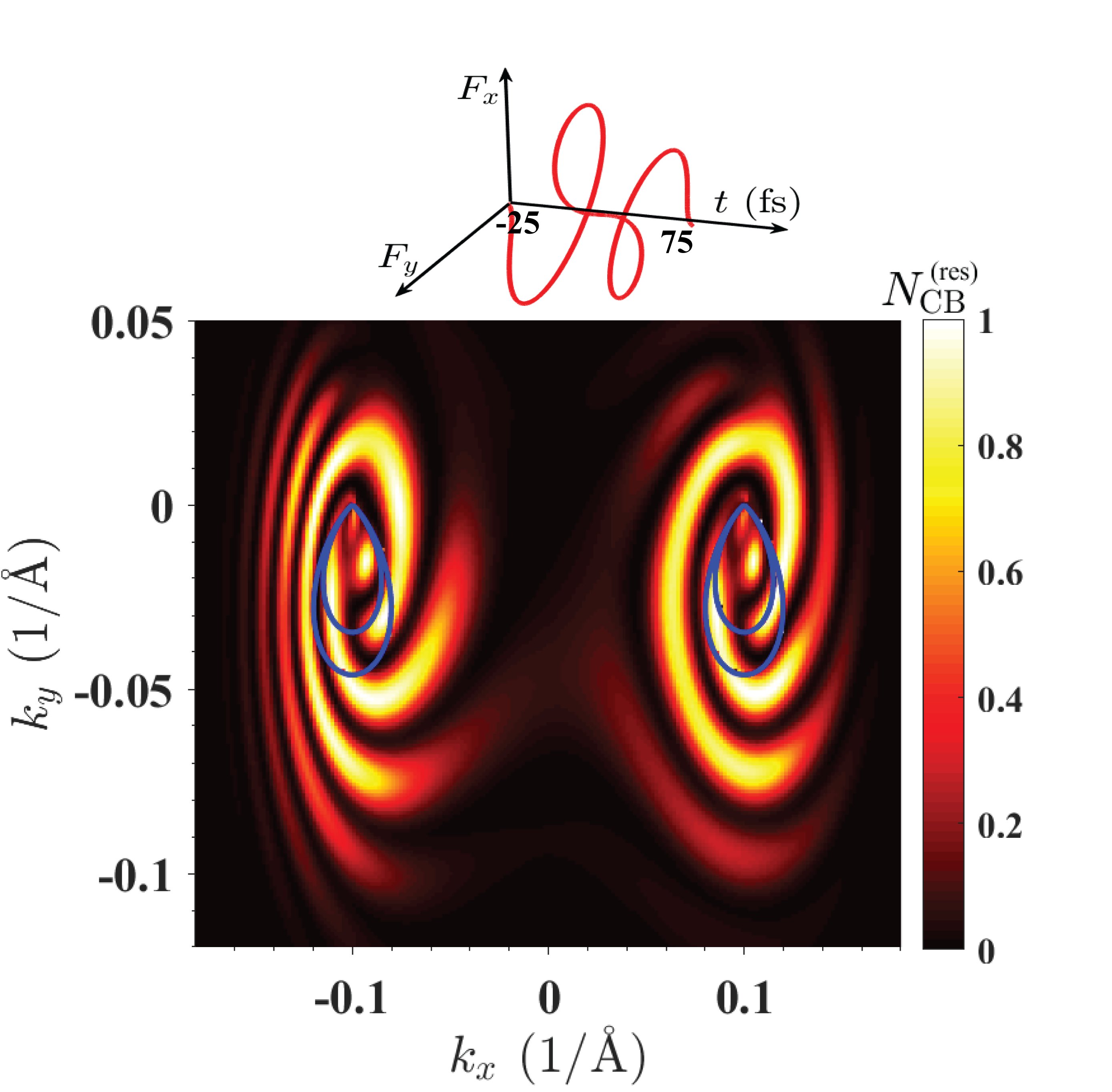}\\\end{center}
\caption{\footnotesize{(Color online) Residual CB population distribution in the reciprocal space as a function of $(k_x,k_y)$ at $k_z=0$, after a two-cycle optical pulse. Here the first cycle of the pulse is right-handed circularly polarized with the amplitude of $F_0=3$ mV/\AA, and the second cycle is left-handed circularly polarized with the amplitude of $0.75F_0$. The separatrix corresponding to two cycles of pulses is shown by a solid blue line.}}
\label{Figure-ccw-cw}
\end{figure}

\begin{figure}
\begin{center}
\includegraphics[width=0.5\textwidth]{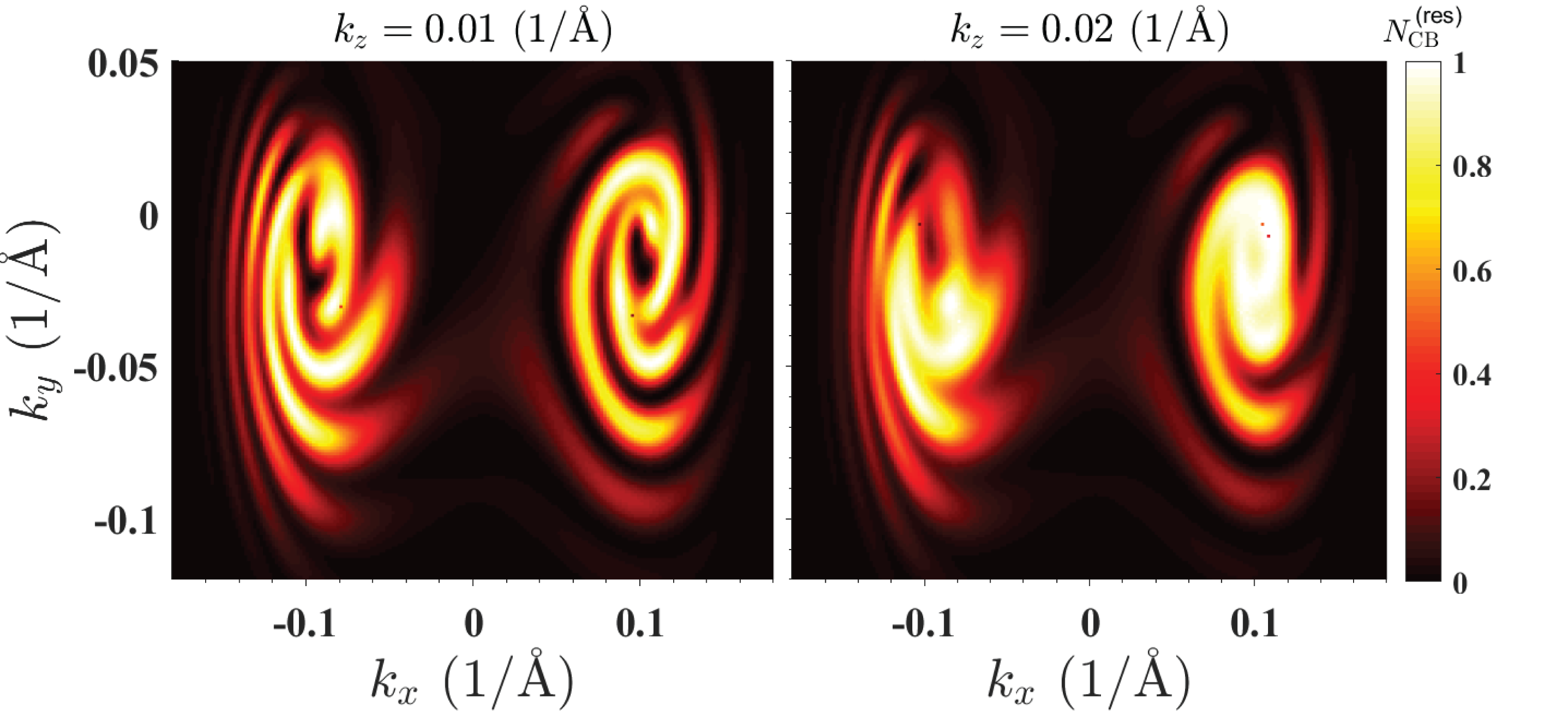}\\\end{center}
\caption{(Color online) Residual CB population as a function of $(k_x,k_y)$ for two non-zero values of $k_z$. The profile of the pulse is the same as the one in Fig.\ref{Figure-ccw-cw}.}
\label{Figure-ccw-cw-qz}
\end{figure}

\begin{figure}[t]
\begin{center}
\includegraphics[width=80mm]{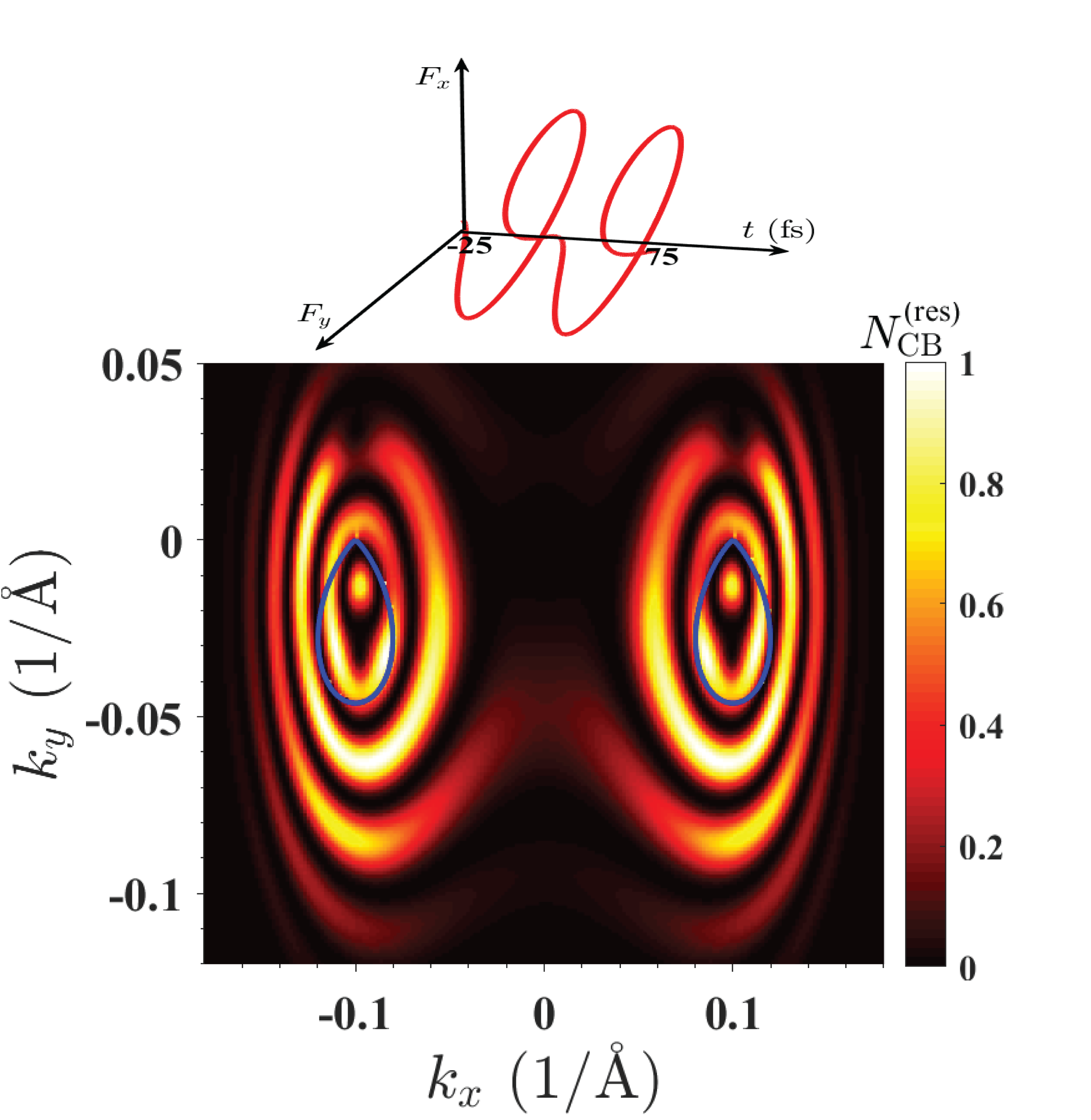}\\
\end{center}
\caption{(Color online) Residual CB population as a function of $(k_x,k_y)$ 
at $k_z=0$ after a two-cycle optical pulse. The two cycles have the same circular polarization. The amplitude of the electric field for both cycles is 
$F_0=3$ mV/\AA. The separatrix is shown be a blue line. The inset shows the 
profile of two-cycle optical pulse.  }   
\label{Figure-ccw-ccw}
\end{figure}

\begin{figure}[t]
\begin{center}
\includegraphics[width=80mm]{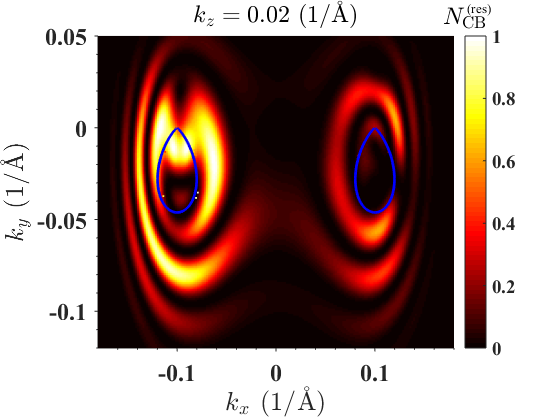}\\
\end{center}
\caption{(Color online) The same as Fig. \ref{Figure-ccw-ccw}, but for $k_z=0.02$ (1/\AA).}   
\label{Figure-ccw-ccw-qz}
\end{figure}

We consider the response  of the Weyl semimetal to a circularly-polarized pulse consisting of two cycles. The pulse is incident normally on the system along $z$ direction and has the following profile 
\begin{align}
F_x(t)&=F_0[-e^{-u^2}(1-2u^2)\mp\alpha e^{-(u-u_0)^2}(1-2(u-u_0)^2)], \nonumber
\\ 
F_y(t)&=2F_0[ue^{-u^2}+\alpha(u-u_0)e^{-(u-u_0)^2}]. 
\label{systemfield}
\end{align}
 Here, $\mp$ sign determines the handedness of the second cycle of the pulse relative to the handedness of the first cycle. Here the "-" sign corresponds to 
the same handedness of two cycles, while the "+" sign corresponds to 
the opposite handedness of two cycles. The amplitude of the first pulse cycle 
is $F_0=3$ mV/\AA, while the amplitude of the second cycle is  $\alpha F_0$, where $\alpha=0.75$ in Figs. \ref{Figure-ccw-cw}-\ref{Figure-ccw-cw-qz} and $\alpha=1$ in Figs. \ref{Figure-ccw-ccw}-\ref{Figure-ccw-ccw-qz}. The duration of a single cycle of the pulse is $\tau$=10 fs and the time interval between the cycles is 
$t_0=50$ fs, where $u_0=t_0/\tau$.

The CB population distribution in the $k_x-k_y$ plane is shown in Fig. (\ref{Figure-ccw-cw}) for $k_z=0$ and for two optical cycles of opposite handedness. 
 The results clearly show that the distribution is highly chiral at both Weyl points and is also characterized by interference fringes. The origin of such 
interference is a double passage by an electron of the region close to the 
Weyl points. Namely, at the end of the first cycle of the pulse, $t\approx 25$ (fs), the CB population distribution has maximum along the corresponding separatrix, see Figs. \ref{Figure_1_pulse} and \ref{Figure_cw}, but does not produce any interference fringes. Then, during the second cycle of the pulse, 
whose circularity is opposite of the circularity of the first cycle, 
electron passes through the Weyl point the second time resulting in the 
interference pattern. Such interference occurs because of highly localized 
nature of the interband dipole coupling, which has sharp maximum at the Weyl points, and because for a two-cycle pulse there are two amplitudes that determine the 
transfer of an electron from the valence band to the conduction band. 
Here the first amplitude corresponds to the electron transfer from VB to CB during the first cycle, while the second amplitude corresponds to the electron transfer 
from VB to CB during the second cycle. The phase, both the dynamic and topological, accumulated between these two transfers determines the interference pattern in the CB population distribution. This interferometer does not need an external reference source and, therefore, is self-referenced.

The results for two-cycle pulse, illustrated in Fig. \ref{Figure-ccw-cw},
show that the CB population distributions are different for two Weyl points, 
while they are the same for one-cycle pulse. Such difference is due to intrinsic 
chirality of electron states at the Weyl points.

The results shown in  Fig. \ref{Figure-ccw-cw} corresponds to $k_z=0$ when 
the interband dipole matrix element is highly localized at the Weyl points and 
there is no topological resonance during the pulse. For the cross-sections 
with $k_z\neq 0$, the electron system behaves similar to gapped graphene with 
well developed topological resonance and broaden interband coupling. 
The corresponding results for a two-cycle pulse is shown in Fig. \ref{Figure-ccw-cw-qz}. The results clearly shown that with increasing $k_z$ the interference 
fringes become smeared. This is mainly related to the fact that for two-cycle pulse, which consists of two cycles with opposite handedness, the topological resonance, for a given Weyl point, occurs only for one of the cycles. As a result, one of the amplitudes that determines CB-VB mixing during the two-cycle pulse becomes small compared to the other one, which finally smears the interference pattern. For example, in Fig. \ref{Figure-ccw-cw-qz}, the first cycle of the pulse has counter-clockwise polarization and it populates the $W$ point, while the second cycle of the pulse has clockwise polarization and populates mainly 
the $W^{\prime}$ point.


The CB population distribution for two-cycle pulse with the same circular polarization and the same amplitude ($\alpha =1$) for two cycles is shown in Figs. \ref{Figure-ccw-ccw} and \ref{Figure-ccw-ccw-qz}. For $k_z=0$, see Fig. \ref{Figure-ccw-ccw},  when the system is similar to pristine graphene, the CB population distributions at two Weyl points are mirror image of each other (with respect to the $k_y-k_z$ plane). The interference fringes in this case are 
mostly parallel to the separatrix, which is shown by blue line in the figure. This is due to the fact that the time between the first and the second passages by an 
electron of the Weyl points is large and is almost the same for all points on a given line parallel to the separatrix. This 
results in strong dephasing and the same interference conditions along such lines. 

For non-zero $k_z$, the system is equivalent to gapped graphene with well pronounced topological resonance. For two-cycle optical pulse with the same handedness for both cycles, the topological resonance is realized only for one of the Weyl points for both cycles. In this case the whole CB population distribution becomes strongly suppressed for one of the Weyl points. This behavior is clearly 
illustrated in Fig. \ref{Figure-ccw-ccw-qz}, where the whole CB population of 
Weyl point $W^\prime $ is suppressed.




\section{Conclusion}

The ultrafast interband electron dynamics in Weyl semimetals  is controlled 
by competition between the dynamic phase and the topological phase, which 
occurs in the field of a circularly polarized femtosecond long optical pulse. 
When these two phases cancels each other the system exhibit the topological 
resonance, which results in large residual conduction band population. 
For Weyl semimetals, for the pulse propagating along, for example, $z$ 
direction, the topological resonance results in predominant CB population 
of the region in the reciprocal space near one of the Weyl points, say $W$, for $k_z<0$ and the region near the other Weyl point, $W^\prime$, for $k_z>0$. Exactly at $k_z=0$ conduction bands at both Weyl points are equally populated. The reason for 
such behavior is that for each cross-section $k_z=const$, the Weyl semimetals behaves as a 2D gapped graphene system with the gap that is proportional to $k_z$.
Since the strength of the topological resonance in gapped graphene system increases with the magnitude of the band gap and manifests itself in predominant population of one of the valleys, then similar features of topological resonance is 
visible in 3D Weyl semimetals. At the same time, the dynamics of Weyl semimetals in circularly polarized pulse can be used to study the properties of topological resonance, i.e., its dependence on the band gap and energy dispersion of the material, and profile and intensity of the optical pulse.

 Furthermore, employing a two-cycle circularly polarized pulse causes the formation of interferogram in the conduction band population distribution in the reciprocal space. Such distribution is also highly chiral for the two Weyl nodes and illustrates the intrinsic chirality of the Weyl points. The interferogram also depends on the strength of the topological resonance and can be used to study 
the properties of the topological resonance in topological materials.  

 

\section{Acknowledgments}
Major funding was provided by Grant No. DE-SC0007043
from the Materials Sciences and Engineering Division of
the Office of the Basic Energy Sciences, Office of Science,
U.S. Department of Energy. Numerical simulations were performed using support by Grant No. DE-FG02-01ER15213
from the Chemical Sciences, Biosciences and Geosciences
Division, Office of Basic Energy Sciences, Office of Science,
U.S. Department of Energy. The work of V.A. was supported
by NSF EFRI NewLAW Grant No. EFMA-17 41691. Support
for F.N. came from MURI Grant No. FA9550-15-1-0037
from the U.S. Air Force Office of Scientific Research.
\bibliography{references}
\bibliographystyle{ieeetr}
\end{document}